\documentclass[12pt,a4paper]{article}
\usepackage[body={16cm,22cm}]{geometry}

\usepackage[utf8]{inputenc}
\usepackage{amsmath, amsthm, amssymb}
\usepackage{enumerate}
\newtheorem{thm}{Theorem}[section]

\newtheorem{tvrzx}[thm]{Proposition}

\newtheorem{lemmax}[thm]{Lemma}
\newenvironment{lemma}{\begin{lemmax}}{\smallskip\end{lemmax}}

\newtheorem{theoremx}[thm]{Theorem}

\theoremstyle{remark}
\newtheorem{remx}[thm]{Remark}

\theoremstyle{definition}
\newtheorem{examplex}[thm]{Example}

\def\R{\mathbb{R}}

\def\<{\langle}
\def\>{\rangle}
\def\~{\widetilde}
\def\^{\wedge}

\def\io{\mathit{i}}

\newcommand{\cffx}[2]{\frac{\partial{#1}}{\partial{x^{#2}}}}

\newcommand{\px}[1]{\~{\partial X}^{#1}}

\begin{document}

\begin{flushright}
\today
\end{flushright}
\vspace{0.7cm}
\begin{center}

\baselineskip=13pt {\Large \bf{$p$-Brane Actions and Higher Roytenberg Brackets}\\}
 \vskip1.3cm
 Branislav Jur\v co$^1$, Peter Schupp$^2$, Jan Vysoký$^3$\\
 \vskip0.6cm
$^{1}$\textit{Mathematical Institute, Faculty of Mathematics and Physics,
Charles University\\ Prague 186 75, Czech Republic, jurco@karlin.mff.cuni.cz}\\
\vskip0.3cm
$^{2}$\textit{Jacobs University Bremen\\ 28759 Bremen, Germany, p.schupp@jacobs-university.de}\\
\vskip0.3cm
$^{3}$\textit{Czech Technical University in Prague\\ Faculty of Nuclear Sciences
and Physical Engineering\\ Prague 115 19, Czech Republic, vysokjan@fjfi.cvut.cz}\\
\vskip0.5cm
\end{center}
\vspace{0.4cm}

\begin{abstract}
Motivated by the quest to understand the analog of non-geometric flux compactification in the context of M-theory, we study higher dimensional analogs of generalized Poisson sigma models and corresponding dual string and $p$-brane models. We find that higher generalizations of the algebraic structures due to Dorfman, Roytenberg and Courant play an important role and establish their relation to Nambu-Poisson structures.
\end{abstract}

{\textit{Keywords:}} Sigma Models, Topological Field Theories, p-Branes, Non-geometric Backgrounds, Flux Compactifications, M-Theory, Bosonic Strings, Nambu-Poisson Structures, Courant-Dorfman Brackets.

\section{Introduction}

To relate ten-dimensional superstring theory to particle physics and cosmology in four-dimensional spacetime, it is necessary to compactify the superfluous dimensions. Introducing fluxes in this context helps to overcome problems of more standard Calabi-Yau compactifications, but at the same time the underlying geometric structures become more general: The notion of a compactifying manifold needs to be relaxed, allowing patching not only by diffeomorphisms but also by more general string symmetry transformations. The resulting non-geometric flux compactifications can appear in the $T$-duals of geometric flux compactifications \cite{Shelton:2005cf,Dabholkar:2005ve}. An example are toroidal compactifications with $R$-fluxes, where non-associative structures arise \cite{Blumenhagen:2011ph}, whose quantization is related to twisted Poisson sigma models \cite{Mylonas:2012pg}.
Poisson sigma models \cite{Ikeda:1993fh,Schaller:1994es} are also at the heart of Kontsevich's approach to deformation quantization \cite{Kontsevich:1997vb}. For a recent review with a comprehensive list of references in the more general context of AKSZ topological field theory, we refer to \cite{Ikeda:2012pv}. See also \cite{dufflu} for an interesting conception of membrane symmetries.

 From a mathematical point of view, it is known that Poisson sigma models are intimately connected to a lot of interesting differential geometry. The fields of Poisson sigma models can be interpreted as Lie algebroid morphisms \cite{liemorphisms} and can be further generalized in terms of generalized (complex) geometry \cite{kotov,kotovstrobl}. It was observed by Alekseev and Strobl in \cite{alekseevstrobl}, that the current algebra of sigma models naturally involves the structures of generalized geometry \cite{Hitchin:2004ut,Gualtieri:2003dx}, such as Dorfman bracket and Dirac structures. This was further developed by Ekstrand and Zabzine in \cite{zabzine} and Bonelli and Zabzine in \cite{zabzinbonelli}. Recently,
D-branes have been identified with Dirac structures \cite{Asakawa:2012px}.
In \cite{halmagyi}, Halmagyi observed that in the Hamiltonian of the Polyakov model, characterized by a $2$-form $B$ and a bivector $\Pi$, appears a more general form of world sheet currents and found their algebra to close under a more general bracket, which he calls a Roytenberg bracket. Finally, in \cite{halmagyi2}, Halmagyi shows that the same bracket appears if one lifts the first order action to a three-manifold using Stokes theorem.

The known string theories as well as supergravity are naturally embedded in eleven-dimensional M-theory, whose building blocks are membranes and five-branes. This motivates the study of higher dimensional analogs of the structures that we have described above. In this article, we would like to go beyond the Courant sigma-model, which is already a higher version of the Poisson sigma-model on an open three-dimensional membrane, but still features a bi-vector field. Generalizing this (twisted) Poisson bi-vector to a $(p+1)$-vector field we face the question how to generalize the Jacobi identity that governs the $p=1$ case. One possibility is to impose the condition of a vanishing Schouten bracket, but that will be non-trivial only for even $p$. Another possibility is to impose the so-called fundamental identity of a Nambu-Poisson structure \cite{Takhtajan:1993vr}. Evidence for the latter choice comes from the study of actions for multiple membranes in M-theory~\cite{Bagger:2006sk}, see \cite{Bagger:2012jb} for a recent review and many references. Local symmetries in M-theory and their relation to generalized geometry were discussed in \cite{Berman:2010is,Berman:2011cg,Berman:2012vc}. For $p=1$, the consistency of the equations of motion of the topological sigma model action implies the Jacobi identity. For $p>1$, the Nambu-Poisson fundamental identity has an algebraic as well as a differential part and it is thus not clear how it could be related to a consistency condition for differential equations of motion. In this article we solve this problem and study the relevant higher algebraic and geometric structures. A suitable higher generalization of Poisson sigma models has recently been proposed by two of us \cite{Jurco:2012yv}: This Nambu-sigma model features a $(p+1)$-dimensional world volume and corresponding higher-order tensor fields on a target manifold. The topological version of the model can also be obtained by an AKSZ construction \cite{bouwjurco}.

This paper is organized as follows:
In section~2 we review the relevant models and compute the Hamiltonian.
In section~3 we use a $(p+1)$-vector $\Pi$ to twist a higher Dorfman bracket and obtain a new Courant bracket like structure, which we call a higher Roytenberg bracket.
In section~4 we discuss the charge algebra of the model and its relation to the higher Roytenberg bracket.
In section~5 we verify the consistency of the topological part of the $p$-brane action. We find that $\Pi$ should satisfy the fundamental identity of a Nambu-Poisson structure (differential as well as algebraic part).
In section~6 we derive the equations of motion of the topological model and find an explicit non-trivial solution.
In section~7 we lift the topological part of the action to a $(p+2)$-dimensional world volume and derive generalized Wess-Zumino terms that involve the structure functions of the higher Roytenberg bracket.
In the appendices we summarize relevant facts about the higher Roytenberg bracket and Nambu Poisson structures.

\section{Nambu sigma model and $p$-brane action}
In this section we review the Nambu sigma model following \cite{Jurco:2012yv,Schupp:2012nq}, compute the corresponding Hamiltonian and remark on the dual $p$-brane action.

Let us consider a $(p+1)$-dimensional world volume $\Sigma$ with a set of local coordinates $(\sigma^{0},\dots,\sigma^{p})$. We assume that $\sigma^{\mu}$ are Cartesian coordinates for a Lorentzian metric $h$ with signature $(-,+,\dots,+)$ on $\Sigma$. Furthermore, we consider an $n$-dimensional target manifold $M$, equipped with a $(p+1)$-vector $\Pi$ and a $(p+1)$-form $B$. We also choose some local coordinates $(y^1,\dots, y^n)$ on $M$. Lower case Latin characters will always correspond to these coordinates. We will use upper case Latin characters to denote strictly ordered multi-indices (mostly $p$-indices), that is $I = (i_1, \dots, i_p)$, where $i_1 < \dots < i_p$.
We will assume that $M$ is equipped with a metric tensor field $G$ with local components $G_{ij}$, and a fiber-wise metric $\~G$ on the vector bundle $\Lambda^{p} TM$ with components $\~G_{IJ}$ in a local section basis $\partial_{I} \equiv \frac{\partial}{\partial y^{i_1}} \^ \cdots \^ \frac{\partial}{\partial y^{i_p}}$. Metric matrices with upper indices denote as usual the corresponding inverses.
For the components of the smooth map $X: \Sigma \rightarrow M$ we will use the following notation: $X^{i} = y^{i}(X)$, $dX^{I} = dX^{i_1} \^ \dots \^ dX^{i_p}$, and $\px{I}=(dX^{I})_{1 \dots p}$ where the latter denotes the  ${1 \dots p}$ component of the world volume form $dX^{I}$.

The ``Nambu-sigma model'' action, as introduced in \cite{Jurco:2012yv,Schupp:2012nq}, is
\begin{multline} \label{def_action}
S[\eta,\~\eta,X] := \int d^{p+1}\sigma \big[ -\frac{1}{2}(G^{-1})^{ij} \eta_{i} \eta_{j} + \frac{1}{2}(\~G^{-1})^{IJ} \~\eta_{I} \~\eta_{J} +
\eta_{i} \partial_{0}X^{i} \\
+ \~\eta_{I} \px{I} - \Pi^{iJ} \eta_{i} \~\eta_{J} - B_{iJ} \partial_{0}X^{i} \px{J}\big]\,,
\end{multline}
where $\eta_{i}, \~\eta_{J} $ are auxiliary fields, which transform under change of local coordinates on $M$ according to their index structure.

The canonical momenta corresponding to the fields $X^{i}$ are
\begin{equation} \label{def_momenta}
P_{i} = \eta_{i} - B_{iJ} \px{J}.
\end{equation}
Starting with the canonical Hamiltonian $H_{can}[X,P,\~\eta] = \int d^{p}\sigma P_{i} \partial_{0}X^{i} - \mathcal{L}(X,P,\~\eta)$ and substituting the Euler-Lagrange equation for $\~\eta_{J}$, we obtain the Hamiltonian\footnote{Note that $\partial_0X^i$ cannot be directly expressed in terms of $P_i$ but it still drops out of $H_{can}$ in the computation, as it should. The construction is robust in the sense that first using the equations of motion for $\~\eta$ and $\eta$ and then constructing the Hamiltonian yields the same result.}
\begin{equation} \label{def_hamiltonian}
H[X,P] = \frac{1}{2}
\int d^{p}\sigma \big[ (G^{-1})^{ij} K_{i} K_{j} + \~G_{IJ} \~K^{I} \~K^{J} \big]\,,
\end{equation}
where
\begin{equation} \label{def_Kd}
K_{i} := \eta_i= P_{i} + B_{iK}\px{K}\,,
\end{equation}
\begin{equation} \label{def_Ku}
\~K^{I} := -\~G^{IJ}\~\eta_{J} = \px{I} - \Pi^{mI} K_{m}\,.
\end{equation}
Here and in the rest of the paper, the integration over $d^{p}\sigma$ means the integration over the space-like coordinates $(\sigma^{1}, \dots, \sigma^{p})$ of $\Sigma$. The Hamiltonian can be conveniently written in matrix notation: The components of the $(p+1)$-vector $\Pi^{iJ}$ form an $n \times \binom{n}{p}$ rectangular matrix $\Pi$ with row index $i$ and column index $J$; similarly for $B$. Likewise, $G$ and $\~G$ are $n \times n$ and $\binom{n}{p} \times \binom{n}{p}$ matrices  corresponding to the metrics $G$ and $\~G$, respectively. Next, we define $(n + \binom{n}{p})$-row column vectors
\[
\mathcal{K} = \left(
\begin{array}{c} K_{i} \\ \~K^{I} \end{array} \right) \quad \text{ and } \quad
\mathcal{V} = \left( \begin{array}{c} P_{i} \\ \px{I} \end{array} \right) \,.
\]
Note that these vectors have the same index structure as coordinate expressions of sections of $T^{\ast}M \oplus \Lambda^{p}TM$. The defining equations (\ref{def_Kd}) and (\ref{def_Ku}) can then be  rewritten as $\mathcal{K} = \mathbf{A} \mathcal{V}$, where
\begin{equation}
\mathbf{A} =
\begin{pmatrix} 1 & 0 \\ -\Pi^T & 1\end{pmatrix} \cdot \begin{pmatrix} 1 & B \\ 0 & 1 \end{pmatrix}
= \begin{pmatrix} \mathbf{1} & B \\
-\Pi^{T} & \mathbf{1} - \Pi^{T} B \end{pmatrix} \,.
\end{equation}
Note that $\mathbf{A}$ can always be inverted, i.e. we can uniquely express the fields $P$ and $\px{}$ using $K$ and $\~K$. We can view $\mathbf{A}$ as the matrix of a linear endomorphism of $T^{\ast}M \oplus \Lambda^{p}TM$. Finally, we can define the matrix $\mathbf{G} = \left( \begin{array}{cc} G^{-1} & 0 \\ 0 & \~G \end{array} \right)$ and view it as the matrix of the fiberwise metric on $T^{\ast}M \oplus \Lambda^{p} TM$. Then, we can rewrite the Hamiltonian (\ref{def_hamiltonian}) as
\begin{equation} \label{hamiltonian_matrix}
H[X,P] = \int d^{p}\sigma [ \mathcal{V}^{T} (\mathbf{A}^{T} \mathbf{G} \mathbf{A}) \mathcal{V} ]   \,.
\end{equation}
Let us note that the matrix $\mathbf{A}^T\mathbf{G}\mathbf{A}$ has a natural interpretation as a (twisted) higher ($p>1$) analog of the generalized metric of  $p=1$ generalized geometry.

If we start again with the action (\ref{def_action}), and integrate out the fields $\~\eta_{J}$ using their equations of motion, we get the action
\begin{equation} \label{action_etaout1}
S[X,\eta] = \int d^{p+1}\sigma \big[ -\frac{1}{2}\eta^{T} G^{-1} \eta - \frac{1}{2} \~K^{T} \~G \~K +
\partial_{0}X^{T} (\eta - B \px{}) \big]\,,
\end{equation}
where $\eta$, $\~\eta$, $K$, $\~K$, $\partial_{0}X$ and $\px{}$ are column vectors defined in the obvious way. We next use the Euler-Lagrange equations to eliminate $\eta$ in (\ref{action_etaout1}) and get
\begin{equation} \label{action_xonly}
S[X] = \int d^{p+1}\sigma \big[ \frac{1}{2} \partial_{0}X^{T} g \,\partial_{0}X - \frac{1}{2} \px{T} \~g \,\px{} -
\partial_{0}X^{T} (B + C)\px{} \big]\,,
\end{equation}
where
\begin{equation} g = (G^{-1} + \Pi \~G \Pi^{T})^{-1}\,, \end{equation}
\begin{equation} \~g = (\~G^{-1} + \Pi^{T}G\Pi)^{-1}\,, \end{equation}
and
\begin{equation} C = -g\Pi\~G = -G\Pi\~g\,. \end{equation}
The action (\ref{action_xonly}) is just the Polyakov-style Howe-Tucker membrane action introduced by Deser-Zumino \cite{Deser:1976rb}, Brink-Di Vecchia-Howe \cite{Brink:1976sc} and Howe-Tucker \cite{Howe:1977hp}  with properly fixed gauge (coordinates on $\Sigma$), see \cite{Jurco:2012yv}. For $p=1$ case, see \cite{Baulieu:2001fi}. 

The background fields $G,\~G,\Pi$ can also be expressed in terms of $g,\~g,C$:
\begin{equation} G = g + C\~g^{-1}C^{T}\,, \end{equation}
\begin{equation} \~G = \~g  + C^{T} g^{-1}C \,, \end{equation}
and
\begin{equation} \Pi = -g^{-1} C \~G^{-1} = -G^{-1} C \~g^{-1}\,. \end{equation}
The relations between $G,\~G,\Pi$ and $g,\~g,C$ are higher $p$-brane
version \cite{Jurco:2012yv} of the well-known open-closed string
relations, cf. also \cite{dufflu}. We can write these relations in
terms of the higher generalized metric $\mathbf{A}^T \mathbf{G}
\mathbf{A}$ as
\[
\mathbf{A}^T \mathbf{G} \mathbf{A} = \mathbf{a}^T \mathbf{g} \mathbf{a} \,, \text{ where }
\mathbf{g} = \begin{pmatrix} g^{-1} & 0 \\ 0 & \tilde g \end{pmatrix} \text{ and }
\mathbf{a} = \begin{pmatrix} 1 & B + C \\ 0 & 1 \end{pmatrix} \,.
\]
The Hamiltonian corresponding to the action (\ref{action_xonly}) features the inverse of the matrix $\mathbf{a}^T \mathbf{g} \mathbf{a}$.

Instead of the $B$-field, it is sometimes more convenient to introduce a $(p+1)$-form $\Phi$ and write $\mathbf{\tilde{A}} = \begin{pmatrix} 1 & \Phi \\ 0 & 1 \end{pmatrix} \cdot \begin{pmatrix} 1 & 0 \\ -\Pi^T & 1\end{pmatrix}$. Redefining $\mathbf{\tilde a} = \begin{pmatrix} 1 & C \\ 0 & 1 \end{pmatrix}$ and equating $\mathbf{\tilde{A}}^T \mathbf{G} \mathbf{\tilde{A}} = \mathbf{\tilde a}^T \mathbf{g} \mathbf{\tilde a}$ provides an alternative derivation of the general open-closed $p$-brane relations of \cite{Jurco:2012yv}. This new approach should also be useful in the context of effective actions for multiple branes ending on branes.
\section{Higher Roytenberg bracket}
In this section we will recall some of the algebraic structures needed in the following.  The name ``Roytenberg bracket'' was introduced by Halmagyi \cite{halmagyi}, since the bracket was originally introduced by Roytenberg in \cite{roytenberg_quasi}. We present a higher analog of this bracket here, which is essentially a higher Dorfman bracket twisted by a $(p+1)$-vector $\Pi$ as well as by a $(p+1)$-form $H$. For further reading on higher Dorfman bracket see e.g. \cite{bisheng} or~\cite{zambon}.

Let $E = TM \oplus \Lambda^{p} T^{\ast}M$. We define a non-degenerate and $C^{\infty}(M)$-bilinear pairing $\<\cdot,\cdot\>: \Gamma(E) \times \Gamma(E) \rightarrow \Omega^{p-1}(M)$ as
\begin{equation} \label{def_pairing} \< V + \xi, W + \eta \> = \io_{V}(\eta) + \io_{W}(\xi), \end{equation}
for vector fields $V,W \in \mathfrak{X}(M)$ and $p$-forms $\xi,\eta \in \Omega^{p}(M)$. We define the anchor map $\rho: E \rightarrow TM$ as the projection onto the first direct summand of $E$, and denote by the same character also the induced map of sections $\rho(V + \xi) = V$. The Dorfman bracket is the $\R$-bilinear bracket on sections $[\cdot,\cdot]_{D}: \Gamma(E) \times \Gamma(E) \rightarrow \Gamma(E)$, defined as
\begin{equation} \label{def_dorfman} [V+\xi,W+\eta]_{D} = [V,W] + \mathcal{L}_{V}(\eta) - \io_{W}(d\xi), \end{equation}
for all $V,W \in \mathfrak{X}(M)$ and $\xi,\eta \in \Omega^{p}(M)$. This bracket is a particular example of a Leibniz algebroid bracket, see \cite{bisheng}. If we define $\mathcal{D}: \Omega^{p-1}(M) \rightarrow \Gamma(E)$ as $\mathcal{D} = j \circ d$, where $j: \Omega^{p}(M) \hookrightarrow \Gamma(E)$ is the inclusion, we have the following properties of Dorfman bracket:
\begin{enumerate}
\item Derivation property:
\begin{equation}
\label{def_jacobi} [e_1,[e_2,e_3]_{D}]_{D} = [[e_1,e_2]_{D},e_3]_{D} + [e_2, [e_1,e_3]_{D}]_{D}\,,
\end{equation}
for all $e_1,e_2,e_3 \in \Gamma(E)$.
\begin{equation}
\label{def_leibniz}
[e_1,fe_2]_D = f[e_1,e_2]_D + (\rho(e_1).f) e_2\,,
\end{equation}
for all $e_1,e_2 \in \Gamma(E)$ and $f \in C^{\infty}(M)$.
\item $\<\cdot,\cdot\>$ is $E$-invariant in the following sense:
\begin{equation} \label{def_Einvariance}
\mathcal{L}_{\rho(e_1)}(\<e_2,e_3\>) = \<[e_1,e_2]_{D},e_3\> + \<e_2, [e_1,e_3]_{D} \>\,,
\end{equation}
for all $e_1,e_2,e_3 \in \Gamma(E)$.
\item Dorfman bracket is skew-symmetric up to ``coboundary'', that is
\begin{equation} \label{def_skewsymcob}
[e,e]_D = \frac{1}{2} \mathcal{D}\<e,e\>\,,
\end{equation}
for all $e \in \Gamma(E)$.
\end{enumerate}
This bracket can be easily modified in two ways:

Firstly, given a  $(p+2)$-form $H \in \Omega^{p+2}(M)$, we can define $H$-twisted higher Dorfman bracket on $E$ as
\begin{equation} \label{def_dorfmanH}
[V+\xi, W+\eta]_{D}^{(H)} = [V,W] + \mathcal{L}_{V}(\eta) - \io_{W}(d\xi) + \io_{W}\io_{V}H.
 \end{equation}

The form $H$ has to be closed, in order to keep the property (\ref{def_jacobi}). All the other properties of higher Dorfman bracket are also valid for the $H$-twisted case.

 Secondly, assume that we have an arbitrary $C^{\infty}(M)$-linear map of sections $\Pi^{\#}: \Omega^{p}(M) \rightarrow \mathfrak{X}(M)$, for example the map induced by a $(p+1)$-vector $\Pi$ on $M$:
\begin{equation} \label{def_pimap} \Pi^{\#}(\xi) = (-1)^{p} \io_{\xi} \Pi = \xi_{K} \Pi^{iK} \partial_{i},\end{equation}
for all $\xi \in \Omega^{p}(M)$. Define new anchor map $\rho: E \rightarrow TM$ as
\begin{equation} \label{def_anchorroyt}
\rho(V+ \xi) = V - \Pi^{\#}(\xi)\,,
\end{equation}
and the ``twisted'' inclusion of $\Omega^{p}(M)$ into $\Gamma(E)$ as
\begin{equation} \label{def_jroyt}
j(\xi) = \xi + \Pi^{\#}(\xi)\,.
\end{equation}
Denote as $pr_2$ the projection onto the second summand  of $E$. Using this notation, one can define new non-degenerate pairing $\<\cdot,\cdot\>_{R}$:
\begin{equation} \label{def_pairingroyt}
\< e_1, e_2 \>_{R} = \io_{\rho(e_1)}(pr_2(e_2)) + \io_{\rho(e_2)}(pr_2(e_1))\,,
\end{equation}
for all $e_1, e_2 \in \Gamma(E)$. Finally, we define the following bracket on $\Gamma(E)$:
\begin{equation} \label{def_roytenberg}
[e_1,e_2]_{R} = [\rho(e_1),\rho(e_2)]
+ j \big( \mathcal{L}_{\rho(e_1)}(pr_2(e_2)) - \io_{\rho(e_2)}(d(pr_2(e_1))) + \io_{\rho(e_2)} \io_{\rho(e_1)} H \big)\,,
\end{equation}
for all $e_1, e_2 \in \Gamma(E)$. We refer to $[\cdot,\cdot]_{R}$ as higher Roytenberg bracket. This bracket together with the anchor map (\ref{def_anchorroyt}) defines again a Leibniz algebroid, i.e., it satisfies (\ref{def_jacobi}) and (\ref{def_leibniz}). More interestingly, it also satisfies (\ref{def_Einvariance}) and (\ref{def_skewsymcob}) with respect to the pairing (\ref{def_pairingroyt}). All of the properties are straightforward to check; see also \cite{bouwjurco}. In appendix~\ref{appendix_1} we present the coordinate form of the higher Roytenberg bracket. For $p=1$ we get exactly the structure functions of \cite{halmagyi2}.

\section{Charge algebra}

In this section we study the algebra of the currents that appear in the Hamiltonian associated to the Nambu-sigma model. We find that the corresponding charge algebra is governed by the higher Roytenberg bracket that we have discussed in the previous section.

Let us return to the Hamiltonian (\ref{def_hamiltonian}). The canonical equal-time Poisson brackets are
\[
\{ X^{i}(\sigma), P_{j}(\sigma') \} = \delta^{i}_{j} \delta(\sigma - \sigma')\,,
\]
where $\sigma, \sigma'$ are the space-like $p$-tuples of world volume coordinates.
We consider the generalized charges
\begin{equation}
\label{def_charge}
Q_{f}(V+ \xi) = \int d^{p}\sigma f(\sigma) [ V^{i} K_{i} + \xi_{J} \~K^{J} ]\,,
\end{equation}
corresponding to the currents $K^{i}$ and $\~K_{J}$ that appear explicitly in the Hamiltonian. Here $V + \xi \in \Gamma(E)$ and $f \in C^{\infty}(\Sigma)$ is a test function.
The appearance of Courant algebroid structures in the current algebra was first observed by Alekseev and Strobl in \cite{alekseevstrobl}
for the Poisson-sigma model, i.e.\ the special case $p=1$. More general observations from the supergeometry point of view were done by Guttenberg in \cite{guttenberg}. Here we will calculate the charge algebra for $p\geq 1$, following the approach of Ekstrand and Zabzine, who integrated the currents to generalized charges. In fact, we shall consider more general charges, involving background fields $\Pi$ and $B$. This can be done in a straightforward manner; however it is easier to use the results of \cite{zabzine}:
With $\~Q_{f}(V+\xi)$ defined as
\begin{equation} \label{def_zabzcharge}
\~Q_{f}(V+\xi) = \int d^{p}\sigma f(\sigma) \big[ V^{i} P_{i} + \xi_{J} \px{J} \big]\,,
\end{equation}
the Poisson bracket is
\begin{multline} \label{zabzine_equation}
\{\~Q_{f}(V+\xi),\~Q_{g}(W+\eta)\} = \\
-\~Q_{fg}([V+\xi,W+\eta]_{D}) - \int d^{p}\sigma g(\sigma) (df \^ X^{\ast}(\<V+\xi,W+\eta\>))_{1 \dots p} \,,
\end{multline}
where $[\cdot,\cdot]_{D}$ is the higher Dorfman bracket (\ref{def_dorfman}) and $\<\cdot,\cdot\>$ is the pairing (\ref{def_pairing}).
We can use this result to find the Poisson brackets for the charges $Q$ as defined in (\ref{def_charge}). The key is the following relation between charges $Q$ and $\tilde Q$:
\begin{equation} \label{charge_correspondence}
Q_{f}(V+\xi) =
\~Q_{f}\big(V - \Pi^{\#}(\xi) + \xi + \io_{V - \Pi^{\#}(\xi)}(B) \big)\,.
\end{equation}
The resulting Poisson bracket of the charges  is
\begin{multline} \label{qcharge_pb}
\{Q_{f}(V+\xi),Q_{g}(W+\eta)\} = \\
-Q_{fg}([V+\xi,W+\eta]_{R})
- \int d^{p}\sigma g(\sigma) (df \^ X^{\ast}(\<V+\xi,W+\eta\>_{R}))_{1 \dots p} \,,
\end{multline}
where $[\cdot,\cdot]_{R}$ is the higher Roytenberg bracket (\ref{def_roytenberg}) and $\<\cdot,\cdot\>_{R}$ is the pairing (\ref{def_pairingroyt}).
The calculation is straightforward but quite lengthy and we omit it here.

Let us note that choosing constant test functions $f=g=1$, one finds that the charge algebra (\ref{qcharge_pb}) closes and it is described by the higher Roytenberg bracket. For the special case $p=1$, this was already observed by Halmagyi \cite{halmagyi}.

Using this result, we can determine conditions for the conservation of such charges. To avoid the anomalous term in (\ref{qcharge_pb}), we shall consider only the charges
\begin{equation} \label{def_totcharge}
Q(V+\xi) := Q_{1}(V+\xi)\,,
\end{equation}
for a constant test function $f=1$. We are interested to obtain conditions on $V + \xi \in \Gamma(E)$, which would guarantee that
\begin{equation} \label{def_chargeconserve}
\{Q(V+\xi),H \} = 0\,,
\end{equation}
where $H$ is the Hamiltonian (\ref{def_hamiltonian}). The left hand side of this condition can be conveniently rewritten using the Leibniz rule for Poisson bracket:
\begin{multline} \label{pb_decomposition}
\{Q(V+\xi),H\} = \\
\frac{1}{2} \{Q(V+\xi),Q_{K_{i}}((G^{-1})^{ij} \partial_{j}) \} + \frac{1}{2} \{Q(V+\xi),Q_{(G^{-1})^{ij} \partial_{j}}(\partial_{i}) \} \\
 + \frac{1}{2} \{Q(V+\xi),Q_{\~K^{I}}(\~G_{IJ} dy^{J})\} + \{Q(V+\xi),Q_{\~G_{IJ} \~K^{J}}(dy^{I}) \} \,.
\end{multline}
Now we can use  (\ref{qcharge_pb}) to carry out the straightforward but tedious calculation that leads to the following result. Let $\mathcal L_W$ be the Lie derivative with respect to the vector field $W = V - \Pi^{\#}(\xi)$.
The following set of conditions ensure that the charge $Q(V+\xi)$ is conserved:
\begin{equation} \label{cond1}
\mathcal{L}_{W}(G)_{ij} = G_{in} \Pi^{nL} \big( W^{m} dB_{mjL} - (d\xi)_{jL} \big) + (i \leftrightarrow j)\,,
\end{equation}
\begin{equation} \label{cond2}
\mathcal{L}_{W}(\~G)_{IJ} = \~G_{IL} \Pi^{nL} \big( W^{m} dB_{mnJ} - (d\xi)_{nJ}\big) + (I \leftrightarrow J)\,,
\end{equation}
\begin{equation} \label{cond3}
\mathcal{L}_{W}(\Pi)^{kI} = \big( \Pi^{kJ} \Pi^{nI} - (\~G^{-1})^{IJ} (G^{-1})^{kn} \big) \big( W^{m} dB_{mnJ} - (d\xi)_{nJ} \big)\,.
\end{equation}
(Here $\~G$ is viewed as a $2p$-times covariant tensor field on $M$.)
Let us observe that there exists a particular simplification of these conditions: Choosing
\begin{equation} \label{cond_dxiiowdb}
d\xi = \io_{W}(dB)\,,
\end{equation}
all terms on the right-hand side vanish and we get a new set of  conditions
\begin{equation} \label{cond_liew}
\mathcal{L}_{W}(G) = \mathcal{L}_{W}(\~G) = \mathcal{L}_{W}(\Pi) = 0\,.
\end{equation}
The special choice (\ref{cond_dxiiowdb}) can be rewritten as
\begin{equation}
\mathcal{L}_{W}(B) = d(\xi - \io_{W}(B))\,.
\end{equation}
The particular solution (\ref{cond_liew}) to the more general conditions  (\ref{cond1}-\ref{cond3})
implies that the image of $V+\xi$ under the anchor map
(\ref{def_anchorroyt}) preserves the background fields $G,\~G,\Pi$ and preserves the $(p+1)$-form field $B$ up to an exact term.

The conditions (\ref{cond1}-\ref{cond3}) have an interesting geometrical meaning. Let $(\cdot,\cdot)$ be the fiberwise metric on $TM \oplus \Lambda^{p} T^{\ast}M$ given by $\mathbf{G}^{-1}$, the inverse of matrix $\mathbf{G}$ appearing in the Hamiltonian (\ref{hamiltonian_matrix}):
\begin{equation} \label{def_genmetric}
(V+\xi,W+\eta) :=
\left( \begin{array}{c} V \\ \xi \end{array} \right)^{T}
\left( \begin{array}{cc} G & 0 \\ 0 & \~G^{-1} \end{array} \right)
\left( \begin{array}{c} W \\ \eta \end{array} \right)\,.
\end{equation}
Let $e = V + \xi \in \Gamma(TM \oplus \Lambda^{p} T^{\ast}M)$. The conditions (\ref{cond1} - \ref{cond3}) are equivalent to the equation
\begin{equation} \label{lem_conservationinvariance}
\rho(e).(e_1,e_2) = ([e,e_1]_{R},e_2) + (e_1,[e,e_2]_{R})\,,
\end{equation}
for all $e_1, e_2 \in \Gamma(TM \oplus \Lambda^{p}T^{\ast}M)$. In the other words, the charge $Q(V+\xi)$ is conserved, if $e = V+\xi$ is a ``Killing section'' of the fiberwise metric $(\cdot,\cdot)$ (\ref{def_genmetric}) with respect to the higher Roytenberg bracket.

\section{Topological model, consistency of constraints}
In this section we examine the topological sigma model, which is obtain from (\ref{def_action}) by setting $G^{-1} = \~G^{-1} = 0$. We will show that algebra of constraints closes on shell and that the constraints are compatible with time evolution. The consistency of the constraints is ensured by the vanishing of certain structure functions of the higher Roytenberg bracket, which in turn is related to the fundamental identity of a Nambu-Poisson structure.

The action has the form
\begin{equation} \label{top_action}
S[\eta,\~\eta,X] := \int d^{p+1}\sigma \big[ \eta_{i} \partial_{0}X^{i} + \~\eta_{I} \px{I} - \Pi^{iJ} \eta_{i}
\~\eta_{J} - B_{iJ}  \partial_{0}X^{i} \px{J}\big]\,.
\end{equation}
The canonical Hamiltonian of this model can be written as
\begin{equation} \label{top_prehamiltonian}
H[X,\~\eta,P] = -
\int d^{p}\sigma \big[ \~\eta_{I} \big(\px{I} - \Pi^{kI} (P_{k} + B_{kJ} \px{J})  \big) \big]\,,
\end{equation}
with canonical momenta $P_k$ as given in (\ref{def_momenta}). Using the notation of (\ref{def_Kd}) and (\ref{def_Ku}), we have
\begin{equation} \label{top_hamiltonian}
H[X,\~\eta,P] = - \int d^{p}\sigma \~\eta_{I} \~K^{I} \,.
\end{equation}
Looking at Lagrange-Euler equation for $\~\eta_{I}$,
we obtain
\begin{equation} \label{constraint}
\~K^{I} = 0\,,
\end{equation}
which should be viewed as a set of constraints, with $\~\eta_{I}$ being the corresponding Lagrange multipliers.
$\~K^{I}$ as well as $H$ can be expressed in terms of the charges (\ref{def_charge}) for special choices of test functions:
\begin{equation} \label{constasQ}
\~K^{I}(\sigma) = Q_{\delta(\sigma-\cdot)}(dy^{I})\,,
\end{equation}
\begin{equation} \label{hamasQ}
H = - Q_{\~\eta_{I}}(dy^{I})\,.
\end{equation}
The constraint algebra and time evolution of constraints can therefore be expressed in terms of the Roytenberg bracket by equation (\ref{qcharge_pb}). In terms of the structure functions of the Roytenberg bracket (cf. appendix) we obtain the following current algebra
\begin{multline} \label{constraints_alegbra}
\{\~ K^I(\sigma),\~ K^J(\sigma')\}= - \delta(\sigma-\sigma')(R^{IJk}K_k + S^{IJ}_K \~K^K)(\sigma') \\
- \big(  d (\delta(\sigma - \cdot)) \^ X^{\ast}(\<dy^{I},dy^{J}\>_{R}) \big)_{1 \dots p}(\sigma')\,.
\end{multline}
It is hence natural to ask for $R$ to vanish. This leads precisely to the condition (\ref{Hnpfieq}) for $\xi = dy^{J}$, $\eta = dy^{I}$ and $H = -dB$. Imposing the condition $R=0$ is thus equivalent to the assumption that $\Pi$ fulfills the differential part of the fundamental identity for a $(-dB)$-twisted Nambu-Poisson tensor\footnote{Note that for $p>1$ the twisting of Nambu-Poisson structures is redundant since it just leads again to an ordinary Nambu-Poisson structure.} and we shall henceforth assume that this is the case.
Even then, there still seems to be a problem with the anomalous last term in (\ref{constraints_alegbra}), since in general the expression
 $\<dy^{J},dy^{I}\>_{R}$ doesn't vanish. To see it let us note that
 vanishing of  $\<dy^{J},dy^{I}\>_{R}$ is equivalent to
\begin{equation} \label{cond_isotropy}
\io_{\Pi^{\#}(dy^{J})}(dy^{I}) + \io_{\Pi^{\#}(dy^{I})}(dy^{J}) = 0\,.
\end{equation}
The anomalous terms can be dealt with using secondary constraints and consistency of these constraints turns out to be ensured by the algebraic part of the fundamental identity for a  Nambu-Poisson tensor.
Indeed, geometrically (\ref{cond_isotropy}) implies that the graph of~$\Pi$, $G_{\Pi} = \{ \xi + \Pi^{\#}(\xi)| \ \xi \in \Omega^{p}(M) \}$, is isotropic with respect to the canonical pairing $\<\cdot,\cdot\>$ (\ref{def_pairing}) on $\Gamma(TM \oplus \Lambda^{p} T^{\ast}M)$. However, as was noticed by Zambon in \cite{zambon}, such (nontrivial) $\Pi$ exists only for $p=1$ and $p = \mbox{dim}M-1$. For $1<p<\mbox{dim}M-1$, we are forced to add the following set of constraints to the system:
\begin{equation} \label{new_constraint}
\chi^{IJ}_{q} \equiv ( X^{\ast} \< dy^{I}, dy^{J}\>_{R} )_{1 \dots \hat{q} \dots p} = 0\,.
\end{equation}
The new constraints (\ref{new_constraint}) do not contain any $P_{m}$'s  and thus they Poisson commute with each other, i.e.
\begin{equation}
\{\chi^{IJ}_{q}(\sigma), \chi^{KL}_r(\sigma')\}=0.
\end{equation}
The Poisson brackets between the new constraints $\chi^{IJ}_{q}$ and the constraints  $\~K^{M}(\sigma')$ are
\begin{multline*}
 \{ \chi^{IJ}_{q}(\sigma), \~K^{M}(\sigma') \} = \big( {S^{MI}}_{K} \chi^{KJ}_{q} + {S^{MJ}}_{K} \chi^{IK}_{q} \big)(\sigma) \delta(\sigma -
\sigma') \\
+ \sum_{\substack{r=1 \\ r \neq q}}^{p} \mbox{sgn}(r,q) \big( X^{\ast}(\io_{\Pi^{\#}(dy^{M})} \< dy^{I}, dy^{J} \>_{R} \>)\big)_{1 \dots \hat{r}
\dots \hat{q} \dots p}(\sigma) \frac{\partial \delta(\sigma' - \cdot)}{\partial \sigma^{r}}(\sigma)\,,
\end{multline*}
where $\mbox{sgn}(r,q)$ is just a sign, irrelevant for the discussion. The first term clearly vanishes for $\chi^{IJ}_{q} = 0$. The second term, in fact, also weakly vanishes (i.e. it vanishes when the constraints equations are used; this is denoted by ``$\approx$'').
To see this, it is sufficient to show that
\begin{equation} \label{condition_vanish}
\big( X^{\ast}( \io_{\Pi^{\#}(dy^{M})} \<dy^{I}, dy^{J}\>_{R}) \big)_{1 \dots \hat{r} \dots \hat{q} \dots p} \approx 0\,,
\end{equation}
Evaluating the left hand side expression at a $p \in
\Sigma$ with $\Pi(X(p)) = 0$, clearly gives zero. If $\Pi(X(p)) \neq 0$, the validity of (\ref{condition_vanish}) can be shown to be a consequence of the following observation made in \cite{hagiwara}
\begin{equation} \label{Hagi}
\<dy^{I},dy^{J}\>_{R}|_{\Lambda^{p-1} \rho(G_{\Pi})} = 0\,,
\end{equation}
where $\rho$ denotes the projection onto the first summand of the graph $G_\Pi$. The reasoning itself is not very illuminating and we skip the details here.\footnote{Alternatively, one can introduce new constraints $\chi^{MIJ}_{rq} := \big( X^{\ast}( \io_{\Pi^{\#}(dy^{M})}
 \<dy^{I}, dy^{J}\>_{R}) \big)_{1 \dots \hat{r} \dots \hat{q} \dots
 p}=0.$ These will obviously Poisson commute with each other and with all
$\chi^{IJ}_q$'s. Hence, we just have to check their Poisson brackets with the $\~K^{I}$'s. Doing this, new anomalous terms proportional to $X^{\ast}(\io_{\Pi^{\#}(dy^{N})} \io_{\Pi^{\#}(dy^{M})} \<dy^{I},dy^{J}\>_{R})$ will appear. We can treat these again as new constraints and repeat the procedure until we arrive at anomalous terms containing $(p-1)$-contractions with $\io_{\Pi^{\#}({sth})}$. By (\ref{Hagi}) this is identically equal to zero. Note that all these auxiliary constraints follow already from the first ones, i.e., from $\chi^{IJ}_{q} = 0$ by the above discussion.}

Since the Hamiltonian is of the form (\ref{top_hamiltonian}), the constraints $\~K^{I}$ and $\chi^{IJ}_q$ are consistent with the dynamics, i.e., they weakly Poisson commute with the Hamiltonian. This follows immediately from the above discussion of the constraints algebra.

To conclude this section, we shall investigate the conservation of charges (\ref{def_totcharge}) with respect to the dynamics governed by the Hamiltonian (\ref{top_hamiltonian}). This is again simple using (\ref{hamasQ}) and (\ref{qcharge_pb}). For the charge $Q(V+\xi)$ to be conserved, one gets the condition
\begin{equation} \label{qcharge_conserve4}\mathcal{L}_{\Pi^{\#}(\~\eta)}(V) = \Pi^{\#}( \io_{\Pi^{\#}(\~\eta)} \io_{V} dB), \end{equation}
where we have introduced the section $\~\eta:= \eta_{J} dy^{J}$ of the pullback bundle $X^{\ast}(\bigwedge^{p} T^{\ast}M)$.
Note that the charge $Q(0+\xi)$ is conserved for arbitrary $\xi \in \Omega^{p}(M)$.

Given the results of this section, we will shall henceforth assume that $\Pi$ is a Nambu-Poisson tensor (which may be twisted in the case $p=1$).

\section{Equations of motion, solution}
In this section we will derive the equations of motion of the topological action (\ref{top_action}) using the Hamiltonian formalism and previous results. Using the natural coordinates associated with every Nambu-Poisson structure (for $p>1$), we will find an explicit solution of these equations. The calculations involve the higher Roytenberg bracket via the charge algebra. They are again quite long, but straightforward and we will mostly just state the results.

Straight from the definition, one can calculate the equations of motion for the $X^{m}$ fields. Indeed, the calculation of
\[ \dot{X}^{m}(\sigma) = \{X^{m}(\sigma),H\}\]
is just an easy application of the Leibniz rule for the Poisson bracket. Of course, among the equations of motion we will find also the constrains $\~K^{I}=0$. The most difficult part comes with the calculation of
\[ \dot{P}_{i}(\sigma) = \{P_{i}(\sigma),H\}\,. \]
This can be done again using (\ref{qcharge_pb}). First, note that
\begin{equation} P_{i} = K_{i} - B_{iL} \big(\Pi^{mL} K_{m} + \~K^{L} \big)\,. \end{equation}
Hence
\[ P_{i}(\sigma) = Q_{\delta(\sigma - \cdot)}\big( \partial_{i} - \Pi^{\#}(\io_{\partial_{i}}B) - \io_{\partial_{i}}B \big)\,. \]
Now, using (\ref{hamasQ}) and (\ref{qcharge_pb}), one gets the following result:
The fields $X^{m}(\sigma), P_{m}(\sigma)$ and $\~\eta_{J}(\sigma)$ of the sigma model defined by action (\ref{top_action}) evolve in accordance with the following set of equations:
\begin{equation} \label{topnsm_eqm1} \px{I} = \Pi^{mI} (P_{m} + B_{mK} \px{K}) \,, \end{equation}
\begin{equation} \label{topnsm_eqm2} \dot{X}^{m} = \Pi^{mJ} \~\eta_{J}\,, \end{equation}
\begin{multline} \label{topnsm_eqm3}
\dot{P}_{m} = - {\Pi^{kJ}}_{,m} P_{k}
- (d\~\eta_{mN} \^ dX^{N})_{1 \dots p} + \Pi^{kJ} B_{kmL} ( d\~\eta_{J} \^ dX^{L} )_{1 \dots p} \\
- \~\eta_{J} \Big( \Pi^{kJ} B_{mL,k} + {\Pi^{kJ}}_{,m} B_{kL} + \sum_{n=1}^{p} {\Pi^{kJ}}_{,l_n} B_{m l_1 \dots k \dots l_p} - \Pi^{kJ}
(dB)_{kmL} \Big) \px{L}   \,.
\end{multline}
In particular, for $B=0$, we get the equations of motion for the untwisted sigma model:
\begin{equation} \label{topfnsm_eqm1} \px{I} = \Pi^{mI} P_{m} \,, \end{equation}
\begin{equation} \label{topfnsm_eqm2} \dot{X}^{m} = \Pi^{mJ} \~\eta_{J}\,, \end{equation}
\begin{equation} \label{topfnsm_eqm3} \dot{P}_{m} = -  \~\eta_{J} {\Pi^{kJ}}_{,m} P_{k}  - (d\~\eta_{mN} \^ dX^{N})_{1 \dots p}\,.
\end{equation}
Now we will show that there always exists a non-trivial solution of the field equations (\ref{topnsm_eqm1}) - (\ref{topnsm_eqm2}).
We will use the natural local coordinates that are associated with every Nambu-Poisson tensor, namely $(x^1, \dots, x^{n})$, such that
\begin{equation} \label{exists0} \Pi = \cffx{}{1} \^ \dots \^ \cffx{}{p+1}\,, \end{equation}
which exist around every point $x \in M$, where $\Pi(x) \neq 0$ (see e.g. \cite{decomposability}). In these coordinates the components of $\Pi$ can be expressed in terms of the Levi-Civita symbol:
\begin{equation} \label{exists1} \Pi^{i_1 \dots i_{p+1}} = \epsilon^{i_1 \dots i_{p+1}}. \end{equation}
This choice of local coordinates  simplifies the equations of motion considerably. We define a $p$-index $[r] = (1, \dots, \hat{r}, \dots p+1)$ and $(p-1)$-index $[p,q] = (1, \dots, \hat{p}, \dots, \hat{q}, \dots, p+1)$. (The hats denote omitted indices.)

The constraints (\ref{new_constraint}) are in these coordinates equivalent to
\[ \frac{\partial X^{m}}{\partial \sigma^{k}} = 0, \]
for $m > p+1$ and $k \in \{1, \dots, p\}$. This is not straightforward to see, one has to use the consequences of (\ref{Hagi}). Furthermore, the equations (\ref{topnsm_eqm2}) impose
\[ \dot{X}^{m} = 0, \]
for $m > p+1$ and we thus get $X^{m} = C^{m}$ for $m>p+1$, where $C^{m} \in \R$ are arbitrary real constants.
One can then easily deduce the following solution of (\ref{topnsm_eqm1} - \ref{topnsm_eqm3}):
\begin{enumerate}[(i.)]
\item For $m \leq p+1$,
\[ X^{m} = f^{m}, \]
where $f^{m} \in C^{\infty}(\Sigma)$ are arbitrary smooth functions on $\Sigma$;
\item for $m > p+1$,
 \[ X^{m} = C^{m} , \]
 where $C^{m} \in \R$ are arbitrary real constants;
\item for $r \leq p+1$,
\[ \~\eta_{[r]} = (-1)^{r+1} \dot{X}^{r}, \]
and if $I \neq [r]$,
\[ \~\eta_{I} = E_{I} ,\]
where $E_{I}$ are arbitrary constants in space-like variables on $\Sigma$.
\item for $r \leq p+1$,
\[ P_{r} = (-1)^{r+1} (1 - B_{1 \dots p+1}) \px{[r]}; \]
\item for $m > p+1$,
\begin{multline*}
P_{m} = \int d\sigma^{0} \Big[ \sum_{k,r=1}^{p+1} \big[ (dB_{km[r]} - B_{m[r],k}) \dot{X}^{k} \px{[r]} \big] \\
 + \sum_{\substack{r,q=1 \\ r \neq q}}^{p+1} \sum_{k=1}^{p+1} B_{km[r,q]}(d\dot{X}^{k} \^ dX^{[r,q]})_{1 \dots p} \Big]\,.
\end{multline*}
\end{enumerate}
Although straightforward, it is actually a lengthy computation to verify that this solution to equations  (\ref{topnsm_eqm1}) and (\ref{topnsm_eqm2}) indeed also solves  the equation (\ref{topnsm_eqm3}).

There is a nice geometrical interpretation of the solutions for $X$: $\Pi$ defines a $(p+1)$-dimensional foliation in $M$, and $(x^1,\dots,x^n)$ are coordinates adapted to this foliation. Hence the fields $X$ are constant in the directions transversal to this foliation.

\section{Generalized Wess-Zumino terms}
In this section, we encounter yet another way how the higher Roytenberg bracket appears in the context of the Nambu sigma model:
Lifting the topological terms of the model to $(p+2)$ dimensions, the structure functions appear as coefficients in the resulting generalized Wess-Zumino terms. This resembles the $p=1$ case, where the generalized Wess-Zumino terms are topological if and only if the associated Roytenberg relations are satisfied.

We shall use the Lagrangian formalism this time and follow essentially the classic approach of Wess, Zumino, and Witten \cite{Wess:1971yu,Witten:1983ar},  adapted to the twisted Poisson sigma model by Halmagyi in \cite{halmagyi2}. 

Define the $p$-forms $A_{i}$ and $1$-forms $\~A_{J}$ as
\[ A_{i} = \eta_{i} d\sigma^{1} \^ \dots \^ d\sigma^{p}, \] 
\[ \~A_{J} = \~\eta_{J} d\sigma^0. \]
Choosing the orientation on $\Sigma$ as $o(\sigma^{0},\sigma^{1},\dots,\sigma^{p}) = +1$ and introducing an auxiliary Minkowski world volume metric, the action (\ref{def_action}) can be rewritten as

\begin{multline} \label{actionA}
S[X,A,\~A] = \int_{\Sigma} -\frac{1}{2} (G^{-1})^{ij} A_{i} \^ \ast{A_{j}} - \frac{1}{2} (\~G^{-1})^{IJ} \~A_{I} \^
\ast{\~A_{J}} \\
 + dX^{i} \^ A_{i} + \~A_{J} \^ dX^{J} - \Pi^{iJ} \~A_{J} \^ A_{i} - X^{\ast}(B)\,,
\end{multline}

Topological part of this action has the form

\begin{multline} \label{actiontop2}
S_{top}[X,A,\~A] = \int_{\Sigma} dX^{i} \^ A_{i} + \~A_{J} \^ dX^{J} - \Pi^{iJ} \~A_{J} \^ A_{i} \\
 + \frac{1}{2} \~A_{I} \^ \~A_{J} \^ M^{IJ} - X^{\ast}(B)\,,
\end{multline}
where we have added a new term $\frac{1}{2} \~A_{I} \^ \~A_{J} \^
M^{IJ}$, which is zero on $\Sigma$\footnote{This term is zero on
$\Sigma = \partial N$; however, we assume an arbitrary extension of
$\~A$ on $N$, hence it is in general non-zero on $N$.} and where
\begin{equation} \label{def_Mthing}
\begin{split}
M^{IJ} &=  \frac{1}{2} X^{\ast}\big(  \io_{\Pi^{\#}(dy^{I})}(dy^{J}) - \io_{\Pi^{\#}(dy^{J})}(dy^{I})\big) \\
&= \frac{1}{2} \sum_{r=1}^{p} (-1)^{r-1} \Pi^{j_rI} ( dX^{j_1} \^ \dots \^ \widehat{dX^{j_r}} \^ \dots \^ dX^{j_p} ) - (I \leftrightarrow J) \,.
\end{split}
\end{equation}
(The hat denotes a factor that is omitted.)

Let us suppose that $\Sigma = \partial{N}$, where $N$ is a smooth $(p+2)$-dimensional manifold. Using Stoke's theorem, we can lift the action to $N$:
\begin{equation} \label{action_lif} S_{top}[X,A,\~A] = \int_{N} d(\mathcal{L})_{top}\,. \end{equation}
\begin{multline*}
 d(\mathcal{L}_{top}) = -(dX^{i} - \Pi^{iJ} \~A_{J}) \^ dA_{i} + d\~A_{J} \^ ( dX^{J} - \Pi^{iJ} A_{i} - M^{JK} \~A_{K} ) \\
 -  {\Pi^{iJ}}_{,k} dX^{k} \^ \~A_{J} \^ A_{i} + \frac{1}{2} \~A_{I} \^ \~A_{J} \^ dM^{IJ} - \frac{p!}{(p+2)!} dB_{klJ} \^ dX^{k} \^
dX^{l} \^ dX^{J} \,.
\end{multline*}
We define new fields $\psi^{i}$ and $\~\psi^{J}$ as
\begin{equation} \label{def_psi1} \psi^{i} = dX^{i} - \Pi^{iJ} \~A_{J}\,, \end{equation}
\begin{equation} \label{def_psi2} \~\psi^{J} = dX^{J} - \Pi^{iJ} A_{i} - M^{JK} \^ \~A_{K}\,. \end{equation}

We observe that
\begin{equation}
\begin{split}
dM^{IJ} &= \frac{1}{2} \sum_{r=1}^{p} {\Pi^{j_rI}}_{,k} dX^{j_1 \dots k \dots j_p} - (I \leftrightarrow J) \\
        &= \frac{1}{2} \sum_{r=1}^{p} {\Pi^{j_rI}}_{,k} \big( \~\psi^{j_1 \dots k \dots j_r} + \Pi^{i j_1 \dots k \dots j_p} A_{i} + M^{j_1
\dots k \dots j_p,K} \^ \~A_{K} \big) - (I \leftrightarrow J)\,.
\end{split}
\end{equation}
Putting the above expression for $dM^{IJ}$ and redefinition of the fields into $d(\mathcal{L}_{top})$, one finds that
\begin{multline}
 d(\mathcal{L}_{top}) = - \psi^{i} \^ dA_{i}  + d\~A_{J} \^ \~\psi^{J}
 + {Q'^{Ji}}_{k} \psi^{k} \^ \~A_{J} \^ A_{i} + \frac{1}{2} {F'_{kl}}^{i} \psi^{k} \^ A_{i} \^ \psi^{l} \\
 - \frac{1}{2} H'_{klJ} \ \psi^{k}
\^ \psi^{l} \^ \~\psi^{J}
 + {D'_{kM}}^{J}  \~A_{M} \^ \~\psi^{J} \^ \psi^{k}\\
 - \frac{1}{2} {S'^{LM}}_{J} \~A_{L} \^ \~A_{M} \^ \~\psi^{J} -\frac{1}{2}
R'^{LJi} \~A_{L} \^ \~A_{J} \^ A_{i} \\
 - \Big( \frac{1}{2} H'_{klL} \psi^{k} \^ \psi^{l} + {D'_{lL}}^{I} \~A_{I} \^ \psi^{l} + \frac{1}{2} {S'^{IJ}}_{L}
\~A_{I} \^ \~A_{J} \Big) \^ \~A_{N} \^ M^{NL}\,,
\end{multline}
where $Q',F',H',D',S',R'$ are structure functions of skew-symmetric version of the higher Roytenberg bracket (see appendix \ref{appendix_1}) corresponding to a re-scaled 3-form flux
\[
H_{jlK} = \frac{1}{(p+1)(p+2)} (dB)_{jlK} \,.
\]

\section{Conclusion}

In this article, we have studied higher dimensional analogs of generalized Poisson sigma models and the corresponding dual string and $p$-brane models. In this context, we have found that higher algebraic structures related to a generalization of the Roytenberg bracket play an important role and that Nambu-Poisson structures are the appropriate $p>1$ generalization of the  Poisson structures that are relevant for the $p=1$ case.

Let us summarize the main results:
By a Legendre transformation, we have obtained the Hamiltonian corresponding to the Nambu sigma model that had been introduced in \cite{Jurco:2012yv} and identified as a dual to the gauge-fixed Polyakov-style Howe-Tucker $p$-brane action. The resulting quadratic form can be viewed as higher-dimensional analog of a generalized metric (see e.g. \cite{zabzinelectures}).
Starting with the definition of a twisted higher Dorfman bracket (see \cite{bisheng}) and using a $(p+1)$-vector $\Pi$, we have further twisted this structure and have obtained a new Courant bracket like structure, which we call a higher Roytenberg bracket. Its $p=1$ version was originally introduced by Roytenberg in \cite{roytenberg_quasi}. We define a higher analog in coordinate-free intrinsic form, such that its properties, which resemble that of higher Dorfman brackets, can be easily verified. The algebraic structures related to this new bracket play a fundamental role throughout this article.
Next, we have defined generalized charges for the model, with a complicated structure that is parameterized by sections of the vector bundle $TM \oplus \Lambda^{p} T^{\ast}M$. We have found that we can use previous results of Ekstrand and Zabzine \cite{zabzine} to calculate the world sheet algebra of the charges. It turns out that the Poisson bracket of the charges closes under the higher Roytenberg bracket up to an anomaly. This anomaly vanishes if one restricts to some isotropic subbundle $TM \oplus \Lambda^{p} T^{\ast}M$ with respect to a twisted pairing $\<\cdot,\cdot\>_{R}$. One can further find the parameterizing sections of the charges, such that they are conserved under time evolution. We have been let to a set of partial differential equations that generalize the ones found by Halmagyi in \cite{halmagyi}. The equations have an interesting geometrical interpretation: They constitute Killing equations with respect to a certain fiber-wise metric.
The topological part of the $p$-brane action turns out to be a system with constraints, as expected. We have analyzed the consistency of these constraints under time evolution and with the constraint algebra itself. The constraints can be written in the terms of the generalized charges that we have introduced in this article and the calculation of their Poisson bracket can be carried out using the higher Roytenberg bracket. Consistency under time evolution forces certain structure functions of the higher Roytenberg bracket to vanish, which is equivalent to the differential part of the fundamental identity satisfied by a Nambu-Poisson tensor. However, an anomalous term remains in the Poisson bracket, which can be dealt with using secondary constraints for the model. We have shown that these secondary constraints are compatible with time evolution, provided that the algebraic part of the fundamental identity of a Nambu-Poisson structure also holds. It is thus natural to consider the background $(p+1)$-vector field $\Pi$ to be a Nambu-Poisson structure.
We have derive explicit expressions for the equations of motion of the topological model, using once more results for the charge algebra. This has been possible, since the canonical momenta $P_{m}$ can be rewritten in the terms of generalized charges. Using  special coordinates, whose existence is guaranteed locally for any Nambu-Poisson structure, we have been able to simplify the equations of motion and find an explicit non-trivial solution. This is similar to the use of Darboux-Weinstein coordinates in the case of Poisson sigma models.
Finally, we have present the analog of the calculation of Halmagyi in \cite{halmagyi2}: We have lifted the topological part of the action to a $(p+2)$-dimensional world volume $N$, such that $\Sigma = \partial N$, using Stoke's theorem. After some redefinitions of the fields, the resulting Lagrangian density (generalized Wess-Zumino terms) incorporates the fields coupled to new background fields, which are the structure functions of the skew-symmetric version of the higher Roytenberg bracket that we have introduce in this paper. The generalized Wess-Zumino terms are topological if and only if the higher Roytenberg relations are satisfied (see appendix \ref{appendix_1}).

Studying the consistency of the topological model, one is let to a set of constraints that are usually understood as constraints on the embedding fields $X$ and eventually imply conditions on the multi-vector $\Pi$, but that can also be interpreted as constraints on the auxiliary fields $\eta$ and $\~\eta$. This was already observed by Halmagyi  in the case $p=1$ in \cite{halmagyi}. Halmagyi does not further comment on the implication of this observation, but we can in fact now understand this in the present context: The constraints on the auxiliary fields effectively reduce the available dimensionality of target space for the other fields of the model. The multi-vector $\Pi$ is of maximal rank in this subspace. It therefore factorizes and is thus forced to be of Nambu-Poisson type. This is true for $p>1$ and confirms the results that we have obtained in this article using more sophisticated methods. The observation and the conclusion is, however, also valid in the well-studied $p=1$ case: A factorized bi-vector (i.e. $\Pi = V_1 \wedge V_2$ with suitable vector fields $V_1$ and $V_2$) will indeed ensure the consistency of the equations of motion, but this is just a special example of a more general Poisson bi-vector satisfying the Jacobi identity, which also ensures consistency.

\section*{Acknowledgement}
We would like to thank Noriaki Ikeda for discussions and helpful
comments on a preliminary version of the manuscript. Also, it is a
pleasure to thank Peter Bouwknegt for discussions. The research of
B.J. was supported by grant GA\v CR P201/12/G028. The research of
J.V. was supported by Grant Agency of the Czech Technical University
in Prague, grant No. SGS10/295/OHK4/3T/14. We also thank to DAAD (PPP)
and ASCR \& MEYS (Mobility) for supporting our collaboration. 
B.J. thanks H. Bursztyn for hospitality at IMPA.
\appendix
\section{Higher Roytenberg bracket, structure functions} \label{appendix_1}
Here we summarize the local form of the higher Roytenberg bracket (\ref{def_roytenberg}) twisted by a $(p+2)$-form flux
\[
H = \frac{1}{(p+1)(p+2)} H_{klJ} dX^k \^ dX^l \^ dX^J  \,,
\]
where $dX^J \equiv dX^{j_1} \^ \ldots \^ dX^{j_p}$ and $J = (j_1,\ldots,j_p)$ denotes an ordered multi-index with $j_1 < \ldots < j_p$.

Let $(y^{1}, \dots, y^{n})$ be a set of local coordinates on $M$. Denote $\partial_{k} = \frac{\partial}{\partial y^{k}}$ and $dy^{K} = dy^{k_1} \^ \dots \^ dy^{k_p}$. Then, one has
\begin{equation} [\partial_{k},\partial_{l}]_{R} = {F_{kl}}^{m} \partial_{m} + H_{klL} dy^{L}\,, \end{equation}
\begin{equation} [\partial_{k},dy^{J}]_{R} = {Q_{k}^{m}}^{J} \partial_{m} + {D_{k}^{J}}_{L} dy^{L}\,, \end{equation}
\begin{equation} [dy^{I},dy^{J}]_{R} = {R^{IJm}} \partial_{m} + {S^{IJ}}_{L} dy^{L}\,. \end{equation}
The structure functions have the following form (Roytenberg relations):
\begin{equation} {F_{kl}}^{m} = H_{klJ} \Pi^{mJ}\,, \end{equation}
\begin{equation} {Q_{k}^{m}}^{J} = - {\Pi^{mJ}}_{,k} + H_{lkL} \Pi^{lJ} \Pi^{mL} \,, \end{equation}
\begin{equation} {D_{k}^{J}}_{L} =  H_{lkL} \Pi^{lJ}\,, \end{equation}
\begin{equation} R^{IJm} = \Pi^{nI} {\Pi^{mJ}}_{,n} - \Pi^{nJ} {\Pi^{mI}}_{,n} - \sum_{r=1}^{p} {\Pi^{j_{r}I}}_{,k} \Pi^{m j_1 \dots k \dots j_p} + \Pi^{kI} \Pi^{lJ} \Pi^{mL} H_{klL}\,, \end{equation}
\begin{equation} {S^{IJ}}_{L} = - \sum_{r=1}^{p} {\Pi^{j_rI}}_{,k} \delta^{j_1 \dots k \dots j_p}_{L} + \Pi^{kI} \Pi^{lJ} H_{klL}\,.
\end{equation}
We denote by a prime the structure functions of the skew-symmetrized version of the higher Roytenberg bracket. For example
${S'^{IJ}}_{L} = \frac{1}{2} \left( {S^{IJ}}_{L} - {S^{JI}}_{L} \right)$.

\section{Nambu-Poisson structures}
Here we recall some fundamental properties of Nambu-Poisson structures \cite{Takhtajan:1993vr} as needed in this paper. For details see, e.g., \cite{hagiwara} or \cite{bisheng}.

For any $(p+1)$-vector field $A$ on $M$ we define the induced map $A^{\#}: \Omega^{p}(M) \rightarrow \mathfrak{X}(M)$ as $A^{\#}(\xi) = (-1)^{p} \io_{\xi}A = \xi_{K} A^{iK} \partial_{i}$.

Let $\Pi$ be a $(p+1)$-vector field on $M$. We call $\Pi$ a Nambu-Poisson structure, if
\begin{equation} \label{def_npfi} \mathcal{L}_{\Pi^{\#}(df_1 \^ \dots \^ df_p)}(\Pi) = 0\,, \end{equation}
for all $f_1, \dots, f_p \in C^{\infty}(M)$.

\begin{lemma}
For arbitrary $p \geq 1$ the condition (\ref{def_npfi}) can be stated in the following equivalent ways:
\begin{enumerate}
\item \label{npfieq} The graph $G_{\Pi} = \{ \Pi^{\#}(\xi) + \xi \ | \ \xi \in \Omega^{p}(M) \}$ is closed under the higher Dorfman bracket
    (\ref{def_dorfman});
\item for any $\xi,\eta \in \Omega^{p}(M)$ it holds that
\begin{equation} \label{npfieq1} (\mathcal{L}_{\Pi^{\#}(\xi)}(\Pi))^{\#}(\eta) = - \Pi^{\#}(\io_{\Pi^{\#}(\eta)}(d\xi))\,; \end{equation}
\item let $[\cdot,\cdot]_{\pi}: \Omega^{p}(M) \times \Omega^{p}(M) \rightarrow \Omega^{p}(M)$ be defined as
\begin{equation} \label{def_leibnizpi} [\xi,\eta]_{\pi} := \mathcal{L}_{\Pi^{\#}(\xi)}(\eta) - \io_{\Pi^{\#}(\eta)}(d\xi)\,, \end{equation}
for all $\xi,\eta \in \Omega^{p}(M)$. Then it holds that
\begin{equation} \label{npfieq2} [\Pi^{\#}(\xi),\Pi^{\#}(\eta)] = \Pi^{\#}([\xi,\eta]_{\pi})\,, \end{equation}
for all $\xi, \eta \in \Omega^{p}(M)$; \item for any $\xi \in \Omega^{p}(M)$ it holds that
\begin{equation} \label{npfieq3} \mathcal{L}_{\Pi^{\#}(\xi)}(\Pi) = - \big( \io_{d\xi}(\Pi) \Pi - \frac{1}{p+1} \io_{d\xi}(\Pi \^ \Pi)
\big)\,. \end{equation}
\end{enumerate}
\end{lemma}

There seems to be a natural way to define a twisted Nambu-Poisson structure:
Let $\Pi$ be a $(p+1)$-vector on $M$. Let $H \in \Omega^{p+2}(M)$, such that $dH = 0$. We call $\Pi$ an $H$-twisted Nambu-Poisson structure, if the graph $G_{\Pi}$ of $\Pi$ is closed under $H$-twisted higher Dorfman bracket (\ref{def_dorfmanH}).
Equivalently, a $H$-twisted Nambu-Poisson structure can be defined using the condition
\begin{equation} \label{Hnpfieq}
(\mathcal{L}_{\Pi^{\#}(\xi)}(\Pi))^{\#}(\eta) = - \Pi^{\#}(\io_{\Pi^{\#}(\eta)}(d\xi -
\io_{\Pi^{\#}(\xi)}H))\,,
\end{equation}
for all $\xi,\eta \in \Omega^{p}(M)$.
This definition is correct, however, for $p>1$ there occurs an interesting thing:
The fundamental identity (\ref{def_npfi}) splits into two parts -- one part is a differential identity similar to the Jacobi identity for the Poisson bivector, the other part of the identity is purely algebraic. Interestingly, the algebraic part of fundamental identity ensures the existence of coordinates $(x^{1}, \dots, x^{n})$ around every point $x$ where $\Pi(x) \neq 0$, such that
\begin{equation} \label{decompo} \Pi = \cffx{}{1} \^ \dots \^ \cffx{}{p+1}\,. \end{equation}
For details, see e.g. \cite{decomposability}. Conversely, every decomposable $(p+1)$-vector is Nambu-Poisson. The algebraic part of (\ref{def_npfi}) comes from the fact that (\ref{npfieq1}) is not $C^{\infty}(M)$-linear in $\xi$. If we now consider (\ref{Hnpfieq}), we see that if we add a part that is $C^{\infty}(M)$-linear in $\xi$, the algebraic part of identity will stay untouched. This means that a $\Pi$ satisfying (\ref{Hnpfieq}) is in fact still an ordinary Nambu-Poisson tensor, satisfying (\ref{npfieq1}). The concept of an $H$-twisted Nambu-Poisson tensor is therefore redundant for $p>1$, as has already been noticed in \cite{bouwjurco}.

\bibliography{nambu_sigma}

\providecommand{\href}[2]{#2}\begingroup\raggedright\begin{thebibliography}{10}

\bibitem{Shelton:2005cf}
J.~Shelton, W.~Taylor, and B.~Wecht, {\it {Nongeometric flux
  compactifications}},  {\em JHEP} {\bf 0510} (2005) 085,
  [\href{http://xxx.lanl.gov/abs/hep-th/0508133}{{\tt hep-th/0508133}}].

\bibitem{Dabholkar:2005ve}
A.~Dabholkar and C.~Hull, {\it {Generalised T-duality and non-geometric
  backgrounds}},  {\em JHEP} {\bf 0605} (2006) 009,
  [\href{http://xxx.lanl.gov/abs/hep-th/0512005}{{\tt hep-th/0512005}}].

\bibitem{Blumenhagen:2011ph}
R.~Blumenhagen, A.~Deser, D.~Lust, E.~Plauschinn, and F.~Rennecke, {\it
  {Non-geometric Fluxes, Asymmetric Strings and Nonassociative Geometry}},
  {\em J.Phys.} {\bf A44} (2011) 385401,
  [\href{http://xxx.lanl.gov/abs/1106.0316}{{\tt arXiv:1106.0316}}].

\bibitem{Mylonas:2012pg}
D.~Mylonas, P.~Schupp, and R.~J. Szabo, {\it {Membrane Sigma-Models and
  Quantization of Non-Geometric Flux Backgrounds}},  {\em JHEP} {\bf 1209}
  (2012) 012, [\href{http://xxx.lanl.gov/abs/1207.0926}{{\tt
  arXiv:1207.0926}}].

\bibitem{Ikeda:1993fh}
N.~Ikeda, {\it {Two-dimensional gravity and nonlinear gauge theory}},  {\em
  Annals Phys.} {\bf 235} (1994) 435--464,
  [\href{http://xxx.lanl.gov/abs/hep-th/9312059}{{\tt hep-th/9312059}}].

\bibitem{Schaller:1994es}
P.~Schaller and T.~Strobl, {\it {Poisson structure induced (topological) field
  theories}},  {\em Mod.Phys.Lett.} {\bf A9} (1994) 3129--3136,
  [\href{http://xxx.lanl.gov/abs/hep-th/9405110}{{\tt hep-th/9405110}}].

\bibitem{Kontsevich:1997vb}
M.~Kontsevich, {\it {Deformation quantization of Poisson manifolds. 1.}},  {\em
  Lett.Math.Phys.} {\bf 66} (2003) 157--216,
  [\href{http://xxx.lanl.gov/abs/q-alg/9709040}{{\tt q-alg/9709040}}].

\bibitem{Ikeda:2012pv}
N.~Ikeda, {\it {Lectures on AKSZ Topological Field Theories for Physicists}},
  \href{http://xxx.lanl.gov/abs/1204.3714}{{\tt arXiv:1204.3714}}.

\bibitem{dufflu}
M.~J. Duff and J.~X. Lu, {\it {Duality Rotations in Membrane Theory}},  {\em
  {Nuclear Physics B}} {\bf {347}} ({1990}) {394--419}.

\bibitem{liemorphisms}
M.~Bojowald, A.~Kotov, and T.~Strobl, {\it {Lie algebroid morphisms, Poisson
  sigma models, and off-shell closed gauge symmetries}},  {\em J.Geom.Phys.}
  {\bf 54} (2005) 400--426, [\href{http://xxx.lanl.gov/abs/math/0406445}{{\tt
  math/0406445}}].

\bibitem{kotov}
A.~Kotov and T.~Strobl, {\it {Generalizing Geometry - Algebroids and Sigma
  Models}},  in {\em Handbook of pseudo-Riemannian geometry and supersymmetry
  (ed. by V. Cortes)}.
\newblock European Mathematical Society, {Z\"{u}rich}, 2010.
\newblock \href{http://xxx.lanl.gov/abs/1004.0632}{{\tt arXiv:1004.0632}}.

\bibitem{kotovstrobl}
A.~Kotov, P.~Schaller, and T.~Strobl, {\it {Dirac sigma models}},  {\em
  Commun.Math.Phys.} {\bf 260} (2005) 455--480,
  [\href{http://xxx.lanl.gov/abs/hep-th/0411112}{{\tt hep-th/0411112}}].

\bibitem{alekseevstrobl}
A.~Alekseev and T.~Strobl, {\it {Current algebras and differential geometry}},
  {\em JHEP} {\bf 0503} (2005) 035,
  [\href{http://xxx.lanl.gov/abs/hep-th/0410183}{{\tt hep-th/0410183}}].

\bibitem{Hitchin:2004ut}
N.~Hitchin, {\it {Generalized Calabi-Yau manifolds}},  {\em Quart.J.Math.Oxford
  Ser.} {\bf 54} (2003) 281--308,
  [\href{http://xxx.lanl.gov/abs/math/0209099}{{\tt math/0209099}}].

\bibitem{Gualtieri:2003dx}
M.~Gualtieri, {\it {Generalized complex geometry}},
  \href{http://xxx.lanl.gov/abs/math/0401221}{{\tt math/0401221}}.

\bibitem{zabzine}
J.~Ekstrand and M.~Zabzine, {\it {Courant-like brackets and loop spaces}},
  {\em JHEP} {\bf 1103} (2011) 074,
  [\href{http://xxx.lanl.gov/abs/0903.3215}{{\tt arXiv:0903.3215}}].

\bibitem{zabzinbonelli}
G.~Bonelli and M.~Zabzine, {\it {From current algebras for p-branes to
  topological M-theory}},  {\em JHEP} {\bf 0509} (2005) 015,
  [\href{http://xxx.lanl.gov/abs/hep-th/0507051}{{\tt hep-th/0507051}}].

\bibitem{Asakawa:2012px}
T.~Asakawa, S.~Sasa, and S.~Watamura, {\it {D-branes in Generalized Geometry
  and Dirac-Born-Infeld Action}},  {\em JHEP} {\bf 1210} (2012) 064,
  [\href{http://xxx.lanl.gov/abs/1206.6964}{{\tt arXiv:1206.6964}}].

\bibitem{halmagyi}
N.~Halmagyi, {\it {Non-geometric String Backgrounds and Worldsheet Algebras}},
  {\em JHEP} {\bf 0807} (2008) 137,
  [\href{http://xxx.lanl.gov/abs/0805.4571}{{\tt arXiv:0805.4571}}].

\bibitem{halmagyi2}
N.~Halmagyi, {\it {Non-geometric Backgrounds and the First Order String Sigma
  Model}},  \href{http://xxx.lanl.gov/abs/0906.2891}{{\tt arXiv:0906.2891}}.

\bibitem{Takhtajan:1993vr}
L.~Takhtajan, {\it {On Foundation of the generalized Nambu mechanics}},  {\em
  Commun.Math.Phys.} {\bf 160} (1994) 295--316,
  [\href{http://xxx.lanl.gov/abs/hep-th/9301111}{{\tt hep-th/9301111}}].

\bibitem{Bagger:2006sk}
J.~Bagger and N.~Lambert, {\it {Modeling Multiple M2's}},  {\em Phys.Rev.} {\bf
  D75} (2007) 045020, [\href{http://xxx.lanl.gov/abs/hep-th/0611108}{{\tt
  hep-th/0611108}}].

\bibitem{Bagger:2012jb}
J.~Bagger, N.~Lambert, S.~Mukhi, and C.~Papageorgakis, {\it {Multiple Membranes
  in M-theory}},  \href{http://xxx.lanl.gov/abs/1203.3546}{{\tt
  arXiv:1203.3546}}.

\bibitem{Berman:2010is}
D.~S. Berman and M.~J. Perry, {\it {Generalized Geometry and M theory}},  {\em
  JHEP} {\bf 1106} (2011) 074, [\href{http://xxx.lanl.gov/abs/1008.1763}{{\tt
  arXiv:1008.1763}}].

\bibitem{Berman:2011cg}
D.~S. Berman, H.~Godazgar, M.~Godazgar, and M.~J. Perry, {\it {The Local
  symmetries of M-theory and their formulation in generalised geometry}},  {\em
  JHEP} {\bf 1201} (2012) 012, [\href{http://xxx.lanl.gov/abs/1110.3930}{{\tt
  arXiv:1110.3930}}].

\bibitem{Berman:2012vc}
D.~S. Berman, M.~Cederwall, A.~Kleinschmidt, and D.~C. Thompson, {\it {The
  gauge structure of generalised diffeomorphisms}},  {\em JHEP} {\bf 1301}
  (2013) 064, [\href{http://xxx.lanl.gov/abs/1208.5884}{{\tt
  arXiv:1208.5884}}].

\bibitem{Jurco:2012yv}
B.~Jurco and P.~Schupp, {\it {Nambu-Sigma model and effective membrane
  actions}},  {\em Phys.Lett.} {\bf B713} (2012) 313--316,
  [\href{http://xxx.lanl.gov/abs/1203.2910}{{\tt arXiv:1203.2910}}].

\bibitem{bouwjurco}
P.~Bouwknegt and B.~Jurco, {\it {AKSZ construction of topological open p-brane
  action and Nambu brackets}},  \href{http://xxx.lanl.gov/abs/1110.0134}{{\tt
  arXiv:1110.0134}}.

\bibitem{Schupp:2012nq}
P.~Schupp and B.~Jurco, {\it {Nambu Sigma Model and Branes}},  {\em PoS
  CORFU2011} (2011) 045, [\href{http://xxx.lanl.gov/abs/1205.2595}{{\tt
  arXiv:1205.2595}}].

\bibitem{Deser:1976rb}
S.~Deser and B.~Zumino, {\it {A Complete Action for the Spinning String}},
  {\em Phys.Lett.} {\bf B65} (1976) 369--373.

\bibitem{Brink:1976sc}
L.~Brink, P.~Di~Vecchia, and P.~S. Howe, {\it {A Locally Supersymmetric and
  Reparametrization Invariant Action for the Spinning String}},  {\em
  Phys.Lett.} {\bf B65} (1976) 471--474.

\bibitem{Howe:1977hp}
P.~S. Howe and R.~Tucker, {\it {A Locally Supersymmetric and Reparametrization
  Invariant Action for a Spinning Membrane}},  {\em J.Phys.} {\bf A10} (1977)
  L155--L158.

\bibitem{Baulieu:2001fi}
L.~Baulieu, A.~S. Losev, and N.~A. Nekrasov, {\it {Target space symmetries in
  topological theories. 1.}},  {\em JHEP} {\bf 0202} (2002) 021,
  [\href{http://xxx.lanl.gov/abs/hep-th/0106042}{{\tt hep-th/0106042}}].

\bibitem{roytenberg_quasi}
D.~Roytenberg, {\it {A Note on quasi Lie bialgebroids and twisted Poisson
  manifolds}},  {\em Lett.Math.Phys.} {\bf 61} (2002) 123--137,
  [\href{http://xxx.lanl.gov/abs/math/0112152}{{\tt math/0112152}}].

\bibitem{bisheng}
Y.~Bi and Y.~Sheng, {\it {On higher analogues of Courant algebroids}},  {\em
  Sci.China} {\bf A54} (2011) 437--447.

\bibitem{zambon}
M.~Zambon, {\it {L-infinity algebras and higher analogues of Dirac structures
  and Courant algebroids}},  \href{http://xxx.lanl.gov/abs/1003.1004}{{\tt
  arXiv:1003.1004}}.

\bibitem{guttenberg}
S.~Guttenberg, {\it {Brackets, Sigma Models and Integrability of Generalized
  Complex Structures}},  {\em JHEP} {\bf 0706} (2007) 004,
  [\href{http://xxx.lanl.gov/abs/hep-th/0609015}{{\tt hep-th/0609015}}].

\bibitem{hagiwara}
Y.~Hagiwara, {\it {Nambu-Dirac manifolds}},  {\em J. Phys. A} {\bf 35} (2002)
  1263.

\bibitem{decomposability}
D.~Alekseevsky and P.~Guha, {\it On decomposability of {N}ambu-{P}oisson
  tensor},  {\em Acta Math. Univ. Comenianae} {\bf LXV} (1996), no.~1 1.

\bibitem{Wess:1971yu}
J.~Wess and B.~Zumino, {\it {Consequences of anomalous Ward identities}},  {\em
  Phys.Lett.} {\bf B37} (1971) 95.

\bibitem{Witten:1983ar}
E.~Witten, {\it {Nonabelian Bosonization in Two-Dimensions}},  {\em
  Commun.Math.Phys.} {\bf 92} (1984) 455--472.

\bibitem{zabzinelectures}
M.~Zabzine, {\it {Lectures on Generalized Complex Geometry and Supersymmetry}},
   {\em Archivum Math.} {\bf 42} (2006) 119--146,
  [\href{http://xxx.lanl.gov/abs/hep-th/0605148}{{\tt hep-th/0605148}}].

\end{thebibliography}\endgroup

\end{document}